\documentclass[pre,amssymb,amsmath,floats,twocolumn,showpacs,superscriptaddress]{revtex4}

\usepackage{epsfig}

\begin{document}

\title{Organization of Complex Networks without Multiple Connections}

\author{S. N. Dorogovtsev}
\affiliation{Departamento de F{\'\i}sica da Universidade de Aveiro, 3810-193 Aveiro, Portugal}
\affiliation{A. F. Ioffe Physico-Technical Institute, 194021 St. Petersburg, Russia}
\author{J. F. F. Mendes}
\affiliation{Departamento de F{\'\i}sica da Universidade de Aveiro, 3810-193 Aveiro, Portugal}
\author{A. M. Povolotsky}
\affiliation{Departamento de F{\'\i}sica da Universidade de Aveiro, 3810-193 Aveiro, Portugal}
\affiliation{Joint Institute for Nuclear Research, 141980 Dubna, Russia} 
\author{A. N. Samukhin}
\affiliation{Departamento de F{\'\i}sica da Universidade de Aveiro, 3810-193 Aveiro, Portugal}
\affiliation{A. F. Ioffe Physico-Technical Institute, 194021 St. Petersburg, Russia}

\date{}

\begin{abstract} 

We find a new structural feature of equilibrium complex random networks 
without multiple and self-connections.  
We show that if the number of connections is sufficiently high, these networks contain a core of highly interconnected vertices. 
The number of vertices in this core varies in the range between $\text{const}\, N^{1/2}$ and $\text{const}\, N^{2/3}$, where $N$ is the number of vertices in a network. 
At the birth point of the core, we obtain the size-dependent cut-off of the 
distribution of the number of connections and find that its position differs from earlier estimates.

\end{abstract}


\maketitle

{\em Introduction.}---Real-world networks are based on more complex architectures 
than the classical random graphs 
which have the  
``trivial'' Poisson distribution of connections \cite{sr51,er59}. 
In 
this sense, the real
networks are complex. 
Equilibrium models \cite{bck01,bl02,dms03,pn04} provide one with a convenient tool for studying the 
architectures of the complex networks. 
The problem of the structural organization of an equilibrium random graph without multiple connections is among the basic problems of the statistical mechanics of networks. 
The point is that the theory of an equilibrium random network with multiple connections \cite{bck01,dms03} can be reduced to the basic non-network problem of the distribution of balls among boxes---``the balls--in--boxes model'' or, as it is also called, the backgammon model \cite{bbj99}. In contrast, the problem of the network without multiple connections is irreducible to simple non-network 
problems. 
The difficulty is  
that the introduced constraint complicates the calculation of the partition function of the statistical ensemble of networks. 
Furthermore, the strong difference from the multiple-connection case can be found only for size-dependent quantities, which demands heavy analytical work, while intuitive arguments are unreliable. 
Due to these difficulties, the problem remained unsolved up to now.   
To describe the structure of the network without multiple connections, 
intuitive arguments \cite{bk03} and simulations \cite{bpv04} were used. 
Also, the generation of degree--degree correlations in 
simple models of networks of this kind 
was studied \cite{msz02,pn03,cp04} (degree is the number of 
connections of a vertex).  

\begin{figure}[b]
\epsfxsize=37mm
\epsffile{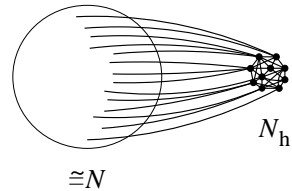}
\caption{ 
Schematic view of the structure of a network without multiple connections if its vertex mean degree exceeds a critical value. The size $N_h(N)$ of the compact core varies in the range between $\text{const}\, N^{1/2}$ and $\text{const}\, N^{2/3}$ vertices. The rest, less connected vertices are shown by a bubble. 
The number of interconnections in the core (as well as the number of outgoing edges) is a finite fraction of all edges. 
In a similar situation, in the multiple network, a finite fraction of all edges are attached to a single hub 
with 
numerous loops of length $1$. 
}
\label{f1}
\end{figure}

In the present Letter, we report the solution of this 
long-standing problem. 
In our study we use a general basic construction of equilibrium networks.  
In the framework of a strict statistical mechanics approach, we describe the equilibrium ensemble of networks without multiple connections where vertices are statistically independent.  
We find that the complex architectures of these networks markedly differ from those for networks with multiple connections. 
We show that at sufficiently high densities of connections---above a critical value, the networks contain a compact core of 
$N_h(N)$, $\text{const}\,N^{1/2}<N_h(N)<\text{const}\,N^{2/3}$, 
highly interconnected vertices [see Fig.~\ref{f1}], where $N$ is the number of vertices in a network. 
In particular, in the networks with slowly decreasing degree distributions, $N_h \sim N^{2/3}$. 
This core is a previously unknown detail of the structure of complex networks. 
We find the form of the degree distribution at the critical point and obtain the position of its size-dependent cut-off. 
In this Letter we present and explain our results and outline the derivation. 
The detailed solution will be published elsewhere.   


{\em The ensembles.}---Following graph theory, we use the standard terms ``multiple graphs'' and ``simple graphs'' for the graphs with and without multiple edges and loops of length $1$, respectively \cite{hbook94}. The equivalent terms are non-Mayer and Mayer graphs, respectively \cite{dms03}. 
The term ``random graph'' means a statistical ensemble of graphs: a set of graphs with their statistical weights. Each graph $g$, a member of a statistical ensemble ${\cal G}$, is described by its adjacency matrix with elements $g_{ij}$. 
In simple graphs, $g_{ii}=0$ and $g_{ij}=0,1$ for $i\neq j$, while  
in multiple graphs, $g_{ii}=0,2,4,\ldots$ and $g_{ij}=0,1,2,\dots$ for $i\neq j$. The vertex degree is $q_i \equiv \sum_{j=1}^N g_{ij}$ (in our ensembles all graphs have equal numbers of vertices, $N$). 
For brevity, we consider sparse networks, where the mean degree is finite in the limit of large $N$ \cite{remark1}. 

In our approach, a random network 
is a 
final 
stationary state of an evolving ensemble. 
The evolution of the ensemble (i.e., transitions between its members---graphs) is due to the processes of reconnection/addition/removal edges 
(see below). 
These processes are governed, e.g., by rules of preferential linking \cite{ba99}, where vertices for linking are selected with probabilities proportional to a function of their degrees---a preference function $f(q)$. 
The resulting network architectures are essentially determined by the form of this function. 
For a wide variety of the preference functions, these equilibrium networks look as follows. (i) If the mean degree $\overline{q}$ is below a critical value $q_c$, the degree distribution $\Pi(q)$ is a rapidly decreasing function. (ii) At the critical point, the degree distribution 
decreases slower than an exponential function. (iii) For $\overline{q}>q_c$, the degree distribution actually coincides with the critical 
one 
plus the edge condensation occurs (in multiple graphs) or the core emerges (in simple graphs).

The members of {\em the grand canonical ensemble} of networks with statistically independent vertices 
are all possible graphs (with a given number of vertices) taken with the following statistical weights \cite{bck01,dms03}: 
\begin{equation} 
P_{GC}(g) = (\lambda N)^{-L(g)}\prod_{i=1}^N \frac{p(q_i)}{2^i(g_{ii}/2)!}\prod_{i<j}\frac{1}{g_{ij}!}
\, ,  
\label{e1}
\end{equation} 
where the parameter $\lambda$ controls the state of the ensemble and is related to the chemical potential (and the fugacity) of edges, 
$L(g) = \frac{1}{2}\sum_{i,j}g_{ij} = \frac{1}{2}\sum_i q_i$ is the number of edges in a graph $g$, and $p(q) = \prod_{r=0}^{q-1}f(r)$. 
The term ``statistically independent vertices'' indicates the specific factorized form of statistical weights (\ref{e1}). 
These weights are the final result of the following assumed network evolution. Two processes take place with equal rates: (i) randomly chosen edges disappear and (ii) new edges connect vertices selected with probability proportional to the product $f(q_i)f(q_j)$ of the preferential functions of their degrees. 
We introduce $\Pi_c(q) \equiv p(q)/q!$, which has a meaning of the degree distribution of the infinite network at the critical point (see below), and its $Z$-transform 
$\Phi(x) = \sum_{q=0}^\infty \Pi_c(q)x^q$. 
Note that this and the next formulas for statistical weights 
of ensembles are valid both for simple and multiple graphs. The partition function $Z_{GC}$ of the grand canonical ensemble is 
the sum of the statistical weights (\ref{e1}) 
over all possible graphs with a given number of vertices. 

The members of {\em the canonical ensemble} of networks with statistically independent vertices 
are all possible graphs with a given number $L$ of edges (and a given number of vertices), ${\cal G}_C$. (Note that ${\cal G}_C$ is a finite set.) 
Their statistical weights are 
\begin{equation} 
P_{C}(g) = N^{-L}\prod_{i=1}^N \frac{p(q_i)}{2^i(g_{ii}/2)!}\prod_{i<j}\frac{1}{g_{ij}!} = \lambda^L P_{GC}(g)
\, .   
\label{e3}
\end{equation} 
These weights are the final result of the following evolution process:  
(i) randomly chosen edges disappear and simultaneously (ii) a new edge connects two vertices selected with probability proportional to the product $f(q_i)f(q_j)$ of the preferential functions of their degrees. 
The partition function $Z_C$ of the canonical ensemble is 
the sum of the statistical weights (\ref{e3}) over graphs belonging to ${\cal G}_C$. 
The partition functions of the ensembles are related to each other: 
\begin{equation} 
Z_{C}(N,L,\{p(q)\}) = \oint \frac{d\lambda}{2\pi i} \lambda^{L-1} Z_{GC}(N,\lambda,\{p(q)\})
\, .  
\label{e5}
\end{equation} 
We have obtained 
the leading asymptotics 
of the partition function of the canonical ensemble of large 
simple graphs in 
a compact 
form. 
The analysis of this function readily leads to the following results. 


\begin{figure}[b]
\epsfxsize=65mm
\epsffile{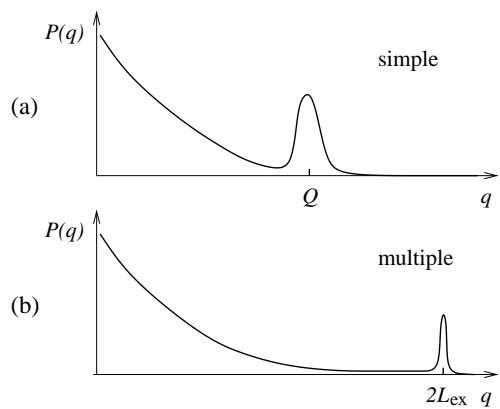}
\caption{ 
Schematic plots of the resulting degree distributions of the equilibrium networks without (a) and with (b) multiple connections in a phase where the mean degree $\overline{q}$ exceeds the critical value $q_c$. 
In simple terms, the degree distributions contain two contributions. 
The first one coincides with the degree distribution at the critical point, $q_c$. 
The second contribution is a peak due to the highly interconnected core vertices of a typical degree 
$Q \sim N/N_h(N)$
in the simple network (a) or a peak due to a single vertex attracting a finite fraction of all connections in the multiple network (b). $L_{\text{ex}}=N(\overline{q}-q_c)$. 
}
\label{f2}
\end{figure}


{\em Results.}---We have found that if the mean degree $\overline{q}$ exceeds a critical 
value $q_c = \sum_{q=0}^\infty q\Pi_c(q)$, the highly interconnected core emerges in the simple graph. This core contains 
\begin{equation} 
N_h
= 
N(\overline{q}-q_c)/Q(\overline{q})
\sim N^{2/3} \ln^{-1/3}\!N
\,   
\label{e5a}
\end{equation} 
vertices with a typical degree $Q \sim N^{1/3}\ln^{1/3}\!N$,  
if 
$f(q)$ is such that 
$\Pi_c(q)$ decreases with $q$ slower than any stretched exponential function. 
Formula (\ref{e5a}) is valid if $N_h$ is large enough. 
A finite fraction of the total number of edges in the network turns out to be inside of the core. Also, a finite fraction of all edges connect the core vertices to the rest weakly connected vertices. 
In contrast, 
in multiple networks with $\overline{q}>q_c$, $N(\overline{q}-q_c)$ edges are connected to a single hub---condensed on it quite similarly to condensation in the backgammon model \cite{bbj99}. A finite fraction of these edges form numerous loops of length $1$. 

The presence of the core is indicated by the form of the resulting degree distribution. 
At $\overline{q}<q_c$, 
the degree distributions are (exponentially) rapidly decreasing for any preference function \cite{bck01,dms03}, and we do not discuss this uninteresting range.  
In contrast, at  $\overline{q} \geq q_c$, the degree distributions have a complex form. 
In particular, if $\Pi_c(q)$ decreases slower than any stretched exponential dependence, the degree distribution of a simple graph is 
\begin{equation} 
\Pi(q) = \Pi_c(q) B^{\!\!\phantom{|_{|_{|_|}}}\!\!\!\!\!(q/Q)-{\displaystyle{\textstyle \frac{1}{2}}}(q/Q)^2} 
\label{e5b}
\end{equation} 
with 
\begin{equation} 
B = \frac{(\overline{q}-q_c)^2}{\pi Q^4\, \Pi_c^2(Q)} 
\,\ln\!\left[ \frac{\overline{q}-q_c}{Q^2\, \Pi_c(Q)} \right]
\, ,   
\label{e5c}
\end{equation} 
where 
the characteristic degree $Q=Q(\overline{q},N)$  
is  
\begin{equation} 
Q
\cong 
\left( \frac{2N\overline{q}^2}{\overline{q}-q_c}\right)^{\!\!1/3}\!
\left| \ln\!\left\{\!\frac{(N\overline{q}^2)^{2/3}}{(\overline{q}-q_c)^{5/3}}\, \Pi_c\!\!\left[\!\left(\frac{N\overline{q}^2}{\overline{q}-q_c}\right)^{1/3}\right]\! \right\}\right|^{1/3}\!\!
.    
\label{e7}
\end{equation} 
Formula (\ref{e5b}) is valid above $q_c$ in the range of parameters where the arguments of the logarithm and $\Pi_c$ in relation (\ref{e7}) are much greater than $1$. 
We find that $Q \sim N^{1/3}\ln^{1/3}N$ and so $B$ in formula (\ref{e5b}) is a (positive) power of $N$. This results in a narrow peak of relative width $\delta q/Q\sim 1/\ln N \ll 1$. 
The resulting degree distribution (\ref{e5b}) is schematically shown in Fig.~\ref{f2}(a). The presence of the peak around $Q$, 
where the area under the peak is $N_h/N$, indicates that in the simple networks, there is a core of $N_h$ highly interconnected vertices. 

For comparison, we obtained the degree distribution of the multiple network   
above $q_c$.
This distribution is schematically shown in Fig.~\ref{f2}(b).  
The distribution contains the ($1-1/N$) contribution of normal vertices and a narrow peak of weight $1/N$ at the degree $2L_{\text{ex}}=2N(\overline{q}-q_c)$, which corresponds to the condensation of the corresponding finite fraction of edges on a single vertex. $L_{\text{ex}}$ is the number of excess edges compared to the number of edges at the critical point $q_c$. 

The form of the degree distributions depends on details of the preference function. In particular, if $f(q) \cong q+1-\gamma$ as $q \to \infty$, the critical state degree distribution is power-law, $\Pi_c(q) \sim q^{-\gamma}$, i.e., with divergent higher moments. If $f(q)$ grows slower, then all the moments of the critical degree distribution converge. 
At the critical point $\overline{q}=q_c$, the degree distribution of the network is equal to $\Pi_c(q)$ modified by finite-size effects.  
At this point, in the simple graphs with with a convergent second moment $m_{2c} = \sum_q q^2 \Pi_c(q) < \infty$, 
\begin{equation} 
\Pi(q,N) = \Pi_c(q) 
\exp\!\left[ -\frac{1}{2N}\!\left(\!\frac{1}{m_{2c}-q_c^2} + \frac{m_{2c}}{q_c^2}\!\right)q^2\right]
. \!  
\label{e8a}
\end{equation} 
At the same point, in the simple networks with power-law $\Pi_c(q) \cong A q^{-\gamma}$, $2<\gamma<3$, which corresponds to $m_{2c} = \infty$, we find 
\begin{equation} 
\Pi(q,N) \!=\! \Pi_c(q) 
\exp\!\left\{\!-\!\left[
\frac{A}{4q_c^2N}\Gamma\!
\left(\!\frac{3-\gamma}{2}\!\right)\right]
^{2/(5-\gamma)\!}
\!\!q^2 \!
\right\}
\!. \! 
\label{e9a}
\end{equation} 
Relations (\ref{e8a}) and (\ref{e9a}) demonstrate a Gaussian size-dependent cutoff of the degree distribution. If $m_{2c} < \infty$ (in particular, if in a scale-free network $\gamma>3$), then the cutoff degree is $q_{\text{cut}} \sim N^{1/2}$. This square-root law fails if $m_{2c} = \infty$, i.e., in particular, when exponent $2<\gamma<3$. In the simple networks with a power-law $\Pi_c(q)$, we have 
\begin{equation} 
\begin{array}{ll}
q_{\text{cut}}(N) \sim N^{1/2}          & \ \ \ \text{if} \ \ \ \gamma>3\, ,  
\\[5pt]
q_{\text{cut}}(N) \sim N^{1/(5-\gamma)} & \ \ \ \text{if} \ \ \ 2<\gamma \leq 3\, .
\end{array}
\label{e10}
\end{equation} 
We emphasize the difference from the multiple networks, where 
the cutoff degree is $q_{\text{cut}} \sim N^{1/2}$ for both the convergent and divergent $m_{2c}$.  

We also described the core in networks with a stretched exponential degree distribution 
$\sim \exp(-\text{const}\,q^\alpha)$, $0<\alpha<1$, which are generated by using the preference function $f(q) = q - \text{const}\,q^\alpha$. In this case, $N_h \propto N^{(2-\alpha)/(3-\alpha)}$, i.e., the exponent of $N_h(N)$ is in the range $(1/2,2/3)$.

{\em Outline of derivations.}---It turns out that 
the grand canonical ensemble exists only 
for 
networks with extremely rapidly decreasing degree distributions. Consequently, we must find the partition function of the canonical ensemble. 
However, as is usual, it is the grand canonical ensemble that admits a convenient analytical consideration. 
So, first we derive the partition function of the grand canonical ensemble of simple graphs and then, by substituting this result into relation (\ref{e5}), obtain the partition function of the canonical ensemble. 

For our purposes, it is convenient to use the inverse Laplace transform of the 
$\Phi(x)$ function:  
$\Psi(z) =\int_{-i\infty }^{+i\infty}\!
e^{-zx}\Phi(x)dx/(2\pi i)$, 
which, at large $z$, coincides with $\Pi_c(z)e^{-z}$ if $\Pi_c(z)$ decays slowly enough. 
In terms of this function, the partition function of the canonical ensembles both of the simple and multiple graphs is  
\begin{equation}
Z_{GC}(N,\lambda) = \int d^N{\bf \!z} \prod_{i=1}^N \Psi(z_i)\exp[f_N(\lambda,{\bf z})]
\, .    
\label{e12}
\end{equation}
For the multiple graph ensemble,  
$f_N(\lambda,{\bf z}) = \sum_{i,j=1}^N z_i z_j/(2N\lambda)$.  
For the ensemble of simple graphs, we use the following idea---the key point of this Letter. 
We substitute the exact function 
\begin{equation}
f_N(\lambda,{\bf z}) = \frac{1}{2} \sum_{i \neq j=1}^N 
\!\ln\! \left(1+ \frac{z_i z_j}{N\lambda}\right) 
\label{e13}
\end{equation}
by the asymptotically exact one: 
\begin{equation}
\frac{1}{2N\lambda}\! \left(\sum_{i=1}^N z_i\!\!\right)^{\!2} \!\!
- \frac{1}{2N\lambda} \sum_{i=1}^N z_i^2 
- \frac{1}{(2N\lambda)^2}\! \left(\sum_{i=1}^N z_i^2\!\right)^{\!2}
\!,     
\label{e13a}
\end{equation}
which allows a convenient analytical consideration. 
The reasons to truncate the expansion are as follows: 
(i) The contributions of simple graphs remain unchanged. 
(ii) The contributions of graphs with single $1$-loops on vertices 
remain zero [note $\sum_{i \neq j}$ in formula (\ref{e13})]. 
(iii) The contributions of graphs with double edge connections remain zero. 
(iv) Graphs with multiple 
$1$-loops 
on a vertex and graphs with triple, quadruple, etc. connections still contribute to the partition function. However, this contribution 
is negligible. 
Indeed, the contribution of any subsequent term in expansion (\ref{e13a}) to the partition function $Z$ may be estimated as $z_0^2/N$ times the contribution of 
the preceding term. 
Here, $z_0$ is the characteristic value of the integration variables $z_i$;  
$z_0 \sim 1$ in the ``normal'' phase, and $z_0 \sim Q \sim N^{1/3}$ in the phase with 
the 
core. 

Using the function (\ref{e13a}) allows us to arrive at the 
following asymptotic 
form of the partition function: 
\begin{eqnarray}
\!\!\!\!\!\!\!\!\!\!\!\!\!\!\!Z_{GC}(N,\lambda) &=& \frac{1}{\pi i} \left(\frac{N\lambda}{2}\right)^{3/2} 
\!\int_{-\infty}^\infty \!dx \int_{-i\infty}^{+i\infty} dy 
\nonumber
\\[7pt]
&& 
\times
e^{-(N\lambda x^2\!/2) + (N\lambda y-1)^2\!/4} \,\Phi_1^N(x,y)
\, ,     
\label{e14}
\end{eqnarray}
where 
\begin{equation}
\Phi_1(x,y) = \int_{-\infty}^\infty \!dz\, e^{xz - yz^2\!/2}\, \Psi(z)
\, .    
\label{e15}
\end{equation}
After substitution of expression (\ref{e14}) into relation (\ref{e5}), the partition function of the canonical ensemble at large $N$ is calculated by using the saddle point approximation with respect to all three integration variables. 
Finally, using the relation 
$
\Pi(q) = N^{-1} \delta\!\ln Z_C(N,L)/\,\delta\!\ln\Pi_c(q)
$ 
leads to the degree distribution.

{\em Interpretation.}---Let us compare our result for the size of the core with characteristic size scales of highly connected graphs with $\sim N$ edges. These networks have two characteristic sizes: 
(i) A fully connected simple graph consists of $N_f(N)\sim N^{1/2}$ vertices. 
(ii) A multiple random graph with, on average, one double connection per vertex consists of $N_1(N) \sim N^{2/3}$ vertices. 
Indeed, the typical degree of a vertex in this network is $Q \sim N/N_1$. 
The probability that a vertex has one double edge must be compared to $1$, i.e., $(Q/N)Q^2 \sim 1$. Here $Q/N$ is the probability that a given pair of edges of a vertex forms a double connections. This probability is multiplied by the total number of edge pairs of the vertex,  
$Q(Q-1)/2 \sim Q^2$. 
So,   
$N^2/N_1^3 \sim 1$ and $N_1 \sim N^{2/3}$. 

Thus, in the networks with slowly decreasing degree distributions, the core is far less densely interconnected than the corresponding fully connected graph. Its size $N_h(N)$ and structure are close to those of the multiple random graph with one double connection per vertex. 
In networks with a stretched exponential degree distribution, the core is more dense, with $N_h(N)$ in the range between those two scales---$N^{1/2}$ and $N^{2/3}$.

{\em Discussion and conclusions.}---A few points should be emphasized. 

(i) 
We considered homogeneous networks. Condensation of edges in inhomogeneous networks, e.g., in networks where the preference function depends on individual properties of vertices and not only on degree, is a different problem \cite{bb01}. 

(ii) 
The cutoff degree $\sim N^{1/(5-\gamma)}$ found at $2<\gamma\leq 3$ differs from the earlier estimate $N^{1/2}$ which was supported 
by heuristic arguments \cite{bk03}. One may show, however, that 
it was actually an upper estimate. 

(iii) 
An empirical researcher usually studies a random network by using only one member of a statistical ensemble. Therefore, he cannot observe the high degree part of the degree distribution, where, in total, less than one vertex occurs. 
For a power-law degree distribution this condition leads to an  
upper 
observable degree 
$\sim N^{1/(\gamma-1)}$ \cite{dms01}. 
This restriction is irrelevant at $2<\gamma\leq 3$, where the cutoff degree $\sim N^{1/(5-\gamma)}$ (\ref{e10}) is smaller than $N^{1/(\gamma-1)}$.  
However, it 
hinders the observation of the $N^{1/2}$ cutoff at $\gamma>3$, where 
$N^{1/(\gamma-1)}$ is smaller than $N^{1/2}$ \cite{bk03,bpv04}. 

(iv) 
In our networks, 
we implemented processes including preferential linking, but it is not a limiting factor. 
Our general results could be equally obtained by using 
network constructions without preferential attachment \cite{bck01,bl02,fdpv04}, where the condensation of edges also occurs.  


In summary, we have described the organization of random networks 
without multiple connections. A new structural feature of complex networks---a highly interconnected compact core---has been revealed.  
The proposed strict statistical mechanics approach may be generalized to networks with various correlations.     

This work was partially supported by projects 
POCTI: FAT/46241/2002, MAT/46176/2002, FIS/61665/2004, and BIA-BCM/62662/2004. 
S.N.D. and J.F.F.M.  
were also supported by project DYSONET-NEST/012911. 
Authors thank  
A.V.~Goltsev for useful discussions.


\begin{thebibliography}{10}
\bibitem{remark1}
Our results may be easily generalized to the case of a network, where the mean degree varies with $N$.
\end{thebibliography}
\end{document}